\documentclass[aps,
prl,
twocolumn,
preprintnumbers,
amsmath,
amssymb,
superscriptaddress,
10pt]{revtex4-1}
\usepackage{graphicx}
\usepackage{bm}
\usepackage[dvipsnames]{xcolor}
\usepackage{color}
\usepackage{braket}
\usepackage{amsmath}
\usepackage{amssymb}
\usepackage{amsfonts}
\usepackage{enumitem}
\usepackage{ulem}
\usepackage[colorlinks]{hyperref}
\usepackage[cp1251]{inputenc}

\usepackage{tempora}					
\usepackage[dvipsnames]{xcolor}			

\begin{document}

\title{Spin noise signatures of the self-induced Larmor precession}

\author{I. I. Ryzhov}
\affiliation{Photonics Department, St.\,Petersburg State University, Peterhof, 198504 St.\,Petersburg, Russia}
\affiliation{Spin Optics Laboratory, St.\,Petersburg State University, Peterhof, 198504 St.\,Petersburg, Russia}

\author{V. O. Kozlov}
\affiliation{Photonics Department, St.\,Petersburg State University, Peterhof, 198504 St.\,Petersburg, Russia}
\affiliation{Spin Optics Laboratory, St.\,Petersburg State University, Peterhof, 198504 St.\,Petersburg, Russia}

\author{N. S. Kuznetsov}
\affiliation{Photonics Department, St.\,Petersburg State University, Peterhof, 198504 St.\,Petersburg, Russia}

\author{I. Yu. Chestnov}
\affiliation{Westlake University, School of Science, 18 Shilongshan Road, Hangzhou 310024, Zhejiang Province, China}
\affiliation{Westlake Institute for Advanced Study, Institute of Natural Sciences, 18 Shilongshan Road, Hangzhou 310024, Zhejiang Province, China}
\affiliation{Vladimir State University, 600000 Vladimir, Russia}

\author{A. V. Kavokin}
\affiliation{Westlake University, School of Science, 18 Shilongshan Road, Hangzhou 310024, Zhejiang Province, China}
\affiliation{Westlake Institute for Advanced Study, Institute of Natural Sciences, 18 Shilongshan Road, Hangzhou 310024, Zhejiang Province, China}
\affiliation{Spin Optics Laboratory, St. Petersburg State University, St. Petersburg 198504, Russia}
\affiliation{Russian Quantum Centre, 100 Novaya St., 143025 Skolkovo, Moscow Region, Russia}

\author{A. Tzimis}
\affiliation{Foundation for Research and Technology-Hellas, Institute of Electronic Structure and Laser, P.O. Box 1527, Heraklion, Crete, 71110, Greece}
\affiliation{Department of Materials Science and Technology, University of Crete, P.O. Box 2208, Heraklion, Crete, 71003, Greece}

\author{Z. Hatzopoulos}
\affiliation{Foundation for Research and Technology-Hellas, Institute of Electronic Structure and Laser, P.O. Box 1527, Heraklion, Crete, 71110, Greece}

\author{P. G. Savvidis}
\affiliation{Westlake University, School of Science, 18 Shilongshan Road, Hangzhou 310024, Zhejiang Province, China}
\affiliation{Westlake Institute for Advanced Study, Institute of Natural Sciences, 18 Shilongshan Road, Hangzhou 310024, Zhejiang Province, China}
\affiliation{Foundation for Research and Technology-Hellas, Institute of Electronic Structure and Laser, P.O. Box 1527, Heraklion, Crete, 71110, Greece}
\affiliation{Department of Materials Science and Technology, University of Crete, P.O. Box 2208, Heraklion, Crete, 71003, Greece}
\affiliation{Department of Nanophotonics and Metamaterials, ITMO University, St.Petersburg, 197101, Russian Federation}

\author{G.~G. Kozlov}
\affiliation{Spin Optics Laboratory, St.\,Petersburg State University, Peterhof, 198504 St.\,Petersburg, Russia}

\author{V.~S. Zapasskii}
\affiliation{Spin Optics Laboratory, St.\,Petersburg State University, Peterhof, 198504 St.\,Petersburg, Russia}

\begin{abstract}
Bose-Einstein condensates of exciton-polaritons are known for their fascinating coherent and polarization
properties. The spin state of the condensate is reflected in polarization of the exciton-polariton emission, with
temporal fluctuations of this polarization being, in general, capable of reflecting quantum statistics of polaritons
in the condensate. To study the polarization properties of optically trapped polariton condensates, we take advantage
of the spin noise spectroscopy technique. The ratio between the noise of ellipticity of the condensate
emission and its polarization plane rotation noise is found to be dependent, in a nontrivial way, on the intensity of continuous wave nonresonant laser pumping. We show that the interplay between the ellipticity and the rotation noise can be
explained in terms of the competition between the self-induced Larmor precession of the condensate pseudospin
and the static polarization anisotropy of the microcavity.
\end{abstract}

\maketitle

\textit{Introduction.} The bosonic condensates of exciton-polari\-tons are unique physical objects exhibiting spontaneous coherence in hybrid driven-dissipative systems. One of their most fascinating properties is an ability to emit a coherent light in a way typical to the conventional lasers but with no need of any population inversion~\cite{yamamoto1996}. The exciton-polaritons arise due to the strong coupling between the semiconductor excitons and the quantized light~\cite{Kavokin2003}. Therefore the macroscopic coherent state of the polariton condensate inherits the spin degree of freedom from both polariton constituents, that is correlated with the polarization of the light emitted by the condensate. In particular, polaritons formed by the coupling of the heavy hole excitons with photons have two possible spin projections on the structure growth axis (either $+1$ or $-1$) which correspond to the right and left circular polarization of the emitted photons~\cite{Shelykh2006}.

Although the exciton-polaritons have been detected in a wide variety of systems \cite{grandjean2008,li2013,dietrich16,pirotta2014,crozier2011} including Fabry--Perot microcavities~\cite{bloch2008,guo13,grandjean2007} and planar waveguides~\cite{jamadi2018}, the most thoroughly investigated system
is a semiconductor microcavity. Polaritons can be excited optically both resonantly and nonresonantly~\cite{bloch2009,savvidis2012}, as well as by the electric current even at room temperature~\cite{grandjean2007,grandjean2008,bhattacharya2014}.

Despite the well developed theoretical basis of exciton-polariton emitters, there is still a number of experimentally unrevealed aspects of their microscopic dynamics. In particular, the effect of self-induced Larmor precession, well known theoretically~\cite{bigenwald2004,kavokin2006} and studied in pulsed and continuous wave (CW) resonant~\cite{roberts2006} experiments, was never observed under nonresonant CW excitation of the condensate. This effect consists in the precession of the Stokes vector of the condensate of exciton-polaritons about an effective magnetic field induced by the ForestGreenpolariton self-interaction within the condensate and oriented along the structure {growth} axis. 
{Being} a peculiar manifestation of the anisotropy of the exchange interaction between excitons in semiconductor quantum wells{, the effect of self-induced Larmor precession was credited by the responsibility for the fast polarization dynamics \cite{roberts2006} of the resonantly excited polariton condensate and for the polarization depinning effect \cite{Read2009,Levrat2010} observed in the pulsed regime of the nonresonant excitation. However, at the CW nonresonant pumping conditions, most of the experimental techniques fail to reveal the presence of the self-induced Larmor precession effect since they provide the information on time-integrated or time-averaged polarization signals.} 
In this context, the spin noise spectroscopy (SNS) technique, widely used in atomic and solid state systems for spin dynamics investigation~\cite{Oestreich-review10,Zapasskii2013,Oestreich-review14} {can be useful. It} was found to be extremely helpful in exciton-polariton condensate studies~\cite{SN-polariton-16}, being informative even in the case when the information seems to be totally lost~\cite{hidden-18}. 

In this paper, we employ the SNS to study the polarization noise properties of the optically trapped polariton condensate. Quite surprisingly, even at almost fully nonpolarized condensate emission, the behaviour of the emission noise demonstrates a strong asymmetry between the ellipticity noise and the polarization plane azimuth noise. We attribute the observed asymmetry to the presence of the self-induced Larmor precession of polarization of the continuously pumped polariton condensate.
This observation paves the way towards application of semiconductor microcavities in optical spintronic devices as proposed in \cite{Shelykh2004}.

\textit{Experimental setup.} To be specific, we study the polarization properties of the polariton condensate formed in the ultra-high-finesse ($Q \approx 31\,000$ corresponding to the polariton lifetime $\tau_p \approx 100$ ps) $5\lambda/2$ microcavity. The cavity is formed by Al$_{0.3}$Ga$_{0.7}$As gap with four sets of three 12 nm GaAs QWs located at the antinodes of the cavity mode. Top (bottom) Bragg mirror is comprised of 45 (50) AlAs/Al$_{0.15}$Ga$_{0.85}$As layers. 

\begin{figure}[h]
\centering
\includegraphics[width=\linewidth]{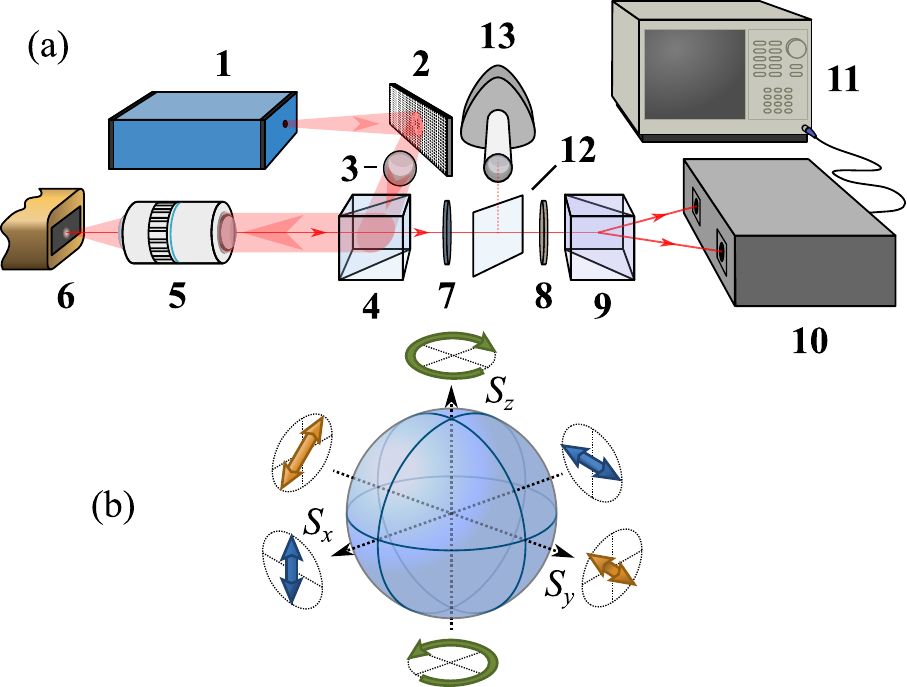}
\caption{(Color online) (a) Schematics of the experimental setup. The description is given in the text. (b) The Poincar\'e sphere with poles corresponding to left- and right-circular polarized light and equator corresponding to various orientations of the polarization plane of a linear polarized light.}\label{setup}
\end{figure}

The polariton condensate is excited by the nonresonant pump of continuous-wave Ti:Sapphire T\&D-scan laser, see Fig.~\ref{setup}~(\textbf{1}). The wavelength of the pump was chosen to match the first next to stop-band reflectivity dip at $\sim$766 nm. The pump laser emission was directed to a FullHD mirror MEMS (DLP)~(\textbf{2}) which serves for the manipulation of the pump beam spatial profile. The reflected image was collimated by a 15 cm lens~(\textbf{3}) and transmitted through a nonpolarizing beamsplitter (\textbf{4}) and 50$\times$~microobjective~(\textbf{5}) to the sample~(\textbf{6}) that was placed in a closed-cycle cold-finger cryostat Montana Cryostation. The emitted light passed in an autocollimation geometry through an optical low-pass filter~(\textbf{7}) to a conventional spin noise detection setup \cite{SN-polariton-16,Zapasskii2013}, which consists of a quarter-wave plate~(\textbf{8}), a polarizing beam splitter~(\textbf{9}), the balanced photoreceiver~(\textbf{10}) and the spectrum analyzer Tektronix RSA5103~(\textbf{11}). The real-space distribution of the condensate emission was sampled by a nonpolarizing plate~(\textbf{12}) and the infrared camera Andor Luca R-604~(\textbf{13}).

\begin{figure}[h]
	\centering
	\includegraphics[width=\linewidth]{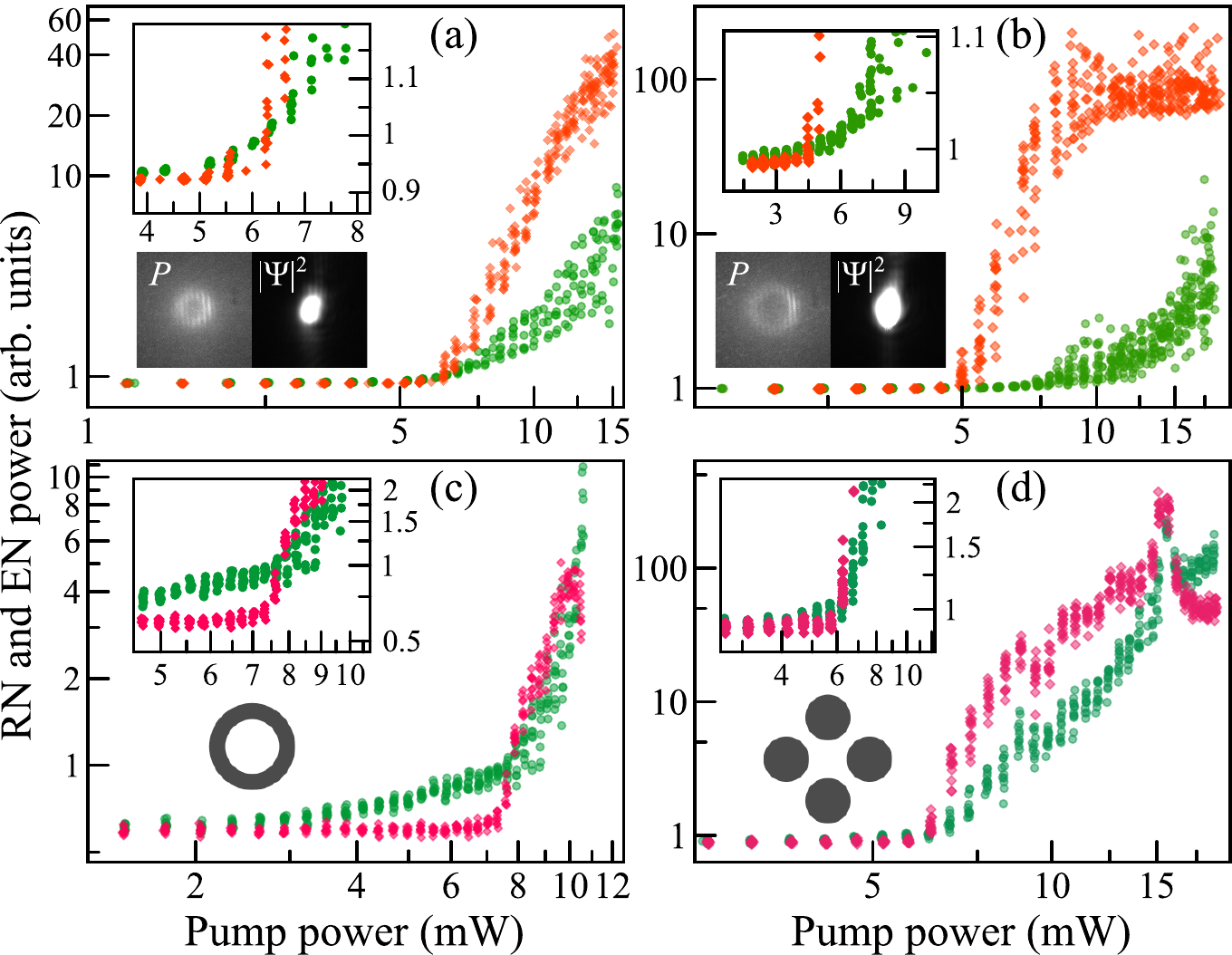}
	\caption{(Color online) The typical experimental dependences of the rotation (green circles) and ellipticity (red rhombi) noise of the exciton-polariton condensate emission versus pump power. Panels (a-d) correspond to different excitation patterns. The pump beam profiles are illustrated by real-space images (top row) or by used DLP patterns (bottom row). The insets show magnified images of the intercrossing part of the dependencies discussed in the text.} \label{meas-examples}
\end{figure}

Polarization measurements require a special care about the spatial and spectral homogeneity of the condensate state. An undesirable fragmentation of the condensate suppresses the polarization signal collected from the whole condensate. In our case, a high degree of homogeneity of the QW interface along with the long polariton lifetime allows for the trapping of the condensate with the pure shape.
Applying the specific spatial profile of the nonresonant pump spot (see the insets in Fig.~\ref{meas-examples}) with the DLP modulator we confine polaritons in an optical trap. The pump creates a cloud of incoherent excitons ~\cite{askitopoulos2013} which push polaritons to the non-excited region due to the exciton repulsion. The polariton confinement leads to the formation of a more pure condensate state than in the case of a Gaussian pump spot profile.

\textit{Results.} The measurements of the polarization degree of the polariton emission (PE) have shown that the condensate was essentially unpolarized at the most of the applied pump beam profiles. The control of the mutual orientation of the quarter-wave plate (\textbf{8}) fast axis and the polarization beam splitter (\textbf{9}) main axis (characterized by angle $\theta$) allows to switch between the detection of the polarization plane azimuth rotation noise, which we address hereafter as the \textit{rotation noise}~(RN), and the \textit{ellipticity noise}~(EN), at $\theta = 0$ and $\theta = 45^{\circ}$, respectively~\cite{SN-polariton-16}.

The measured polarization noise parameters can be related to the Stokes parameters of the polariton condensate emission. The latter, in turn, directly correlates with the condensate pseudospin vector $\mathbf{S}=\frac{1}{2}(\mathbf{\Psi}^\dag\bm{\sigma}\mathbf{\Psi})$,
where $\sigma_{x,y,z}$ are the Pauli matrices and $\mathbf{\Psi}= (\Psi_+, \Psi_-)^{\rm T}$ is the condensate order parameter whose components define two possible projections of the polariton spin on the structure growth axis, see Fig.~1b.
In particular, we take the horizontal and the vertical Stokes polarization (polariton pseudospin) axes coincident with the Cartesian axes of the polarization beam splitter. Then, at $\theta = 0$ the spin noise detection scheme tracks the noise dynamics of the $S_x$ component of the polariton pseudospin. Likewise, if the quarter-wave plate is tilted to $\theta = 45^{\circ}$, the circularly polarized component of emission is converted to the linear polarization, and the balanced scheme detects the intensity variations of circularly polarized components of the emitted light, i.e. the fluctuations of the $S_z$ parameter.

Fixing the shape of the pump spot, we performed several tens of pump intensity scans for both $\theta = 0^{\circ}$ and $\theta = 45^{\circ}$ quarter-wave plate positions. For each value of the pump intensity we collected the integral of the balanced photodiode current noise power in the frequency range of [0...1] MHz.

Although the obtained dependencies (see Fig.~\ref{meas-examples}) {at high pump powers substantially} varied for different excitation beam profiles, in a vast majority of cases, especially for the highly symmetrical pumps, the noise behaviour {around the polariton laser threshold point} follows the same typical scenario. In particular, the rotation noise dominates over the noise of ellipticity at the weak pump, prevalently below lasing threshold (see the insets of Fig.~\ref{meas-examples}). With the increase of the pump intensity, the noise signals of both types progressively grow. However, the EN power growth rate typically exceeds the increment of the RN power. As a result, at some point above the polariton lasing threshold EN~signal becomes greater than the RN. {The
complex behavior of the curves in the high power limit can
arise from the multistability of the trapped condensate state~\cite{Dreismann16} and/or from the loss of strong coupling and the threshold to photon lasing~\cite{SN-polariton-16}. Here we focus at the behaviour of RN and EN signal at low and intermediate pump power, where the system can be described in a single-mode approximation.}

\textit{Discussion.} We interpret the observed behavior of the condensate polarization noise in terms of the interplay between the static polarization anisotropy, which tends to suppress the circular polarization, and the spin-dependent polariton-polariton interactions, which trigger the self-induced Larmor precession of the condensate pseudospin. 
We describe the state of the fluctuating order parameter of the polariton condensate in the presence of the non-resonant pump by the stochastic driven-dissipative Gross-Pitaevskii equation \cite{Read2009}:
\begin{multline}
i \hbar \partial_t \Psi_{\pm} = \left( \alpha_1 {\left| \Psi_{\pm} \right|}^2 + \alpha_2 {\left| \Psi_{\mp} \right|}^2 + \alpha_r n \right) \Psi_{\pm} + \\
+ \frac{i \hbar}{2} (Rn - \gamma_c) \Psi_{\pm} + \frac{\delta_l}{2} \Psi_{\mp} + \sqrt{D} \eta_{\pm}(t),
\label{eq1}
\end{multline}
for the two-component complex order parameter $\boldsymbol{\Psi} = (\Psi_+, \Psi_-)^{\rm T}$, whose indices correspond to polariton $\pm 1$~spin {states}. Coefficients $\alpha_1$, $\alpha_2$
define the interaction strengths between polaritons with the same and the opposite spins, respectively, while 
$\alpha_r$ stands for the interaction of the polaritons from the condensate with the incoherent excitons created by the nonresonant pump; $\gamma_c$ is the polariton lifetime. The quantity $\delta_l$ in the next-to-last term describes the splitting of the linear polarizations that stems from the optical anisotropy of the microcavity {and, in our case, is ascribed to the inevitable mechanical strain of the structure due to the lattice disregistry of the Bragg mirror layers}.

The random fluctuations of the order parameter are accounted for by the last term in Eq.~\eqref{eq1}. Taking into account the high homogeneity of the sample microcavity structure, we attribute the main source of these fluctuations to the scattering from the incoherent reservoir to the condensate state, assuming that every condensate realization is single-mode. Each act of the scattering perturbs the phase of both spin components of the condensate, thus randomizing its polarization.
We simulate $\eta_\pm(t)$ as a Wiener process with the spectral density $D=\langle \eta_i(t)\eta^*_j({t^\prime}) \rangle = \frac{1}{2}\,Rn \, \delta(t-t^\prime)\delta_{ij}$ (see \cite{Read2009}), where $(i,j)=(+,-)$, $n$ corresponds to the incoherent exciton density and $R$ is the scattering rate. Thus the magnitude of the order parameter fluctuations appears to be dependent on the pump power {insofar as the noncondensed exciton population depends on it}.

Since the pumping laser frequency was tuned far above the exciton reservoir, we assume the loss of the memory of the pumping laser polarization, considering the spin-unpolarized reservoir, which integrated density $n$ is governed by the rate equation
\begin{equation}
\partial_t n = P - \gamma_r n - Rn \left( |\Psi_+|^2 + |\Psi_-|^2 \right),
\label{eq2}
\end{equation}
where $P$ defines the pump intensity and $\gamma_r$ is a reservoir population relaxation parameter.

Note that for simplicity we neglect the spatial distributions of $\boldsymbol{\Psi}$ and $n$. This approach allows us to describe the typical behaviour of the PE noise which is independent on the optical trap shape.
\begin{figure*}[ht!]
	\centering
	\includegraphics*[width=0.9\linewidth]{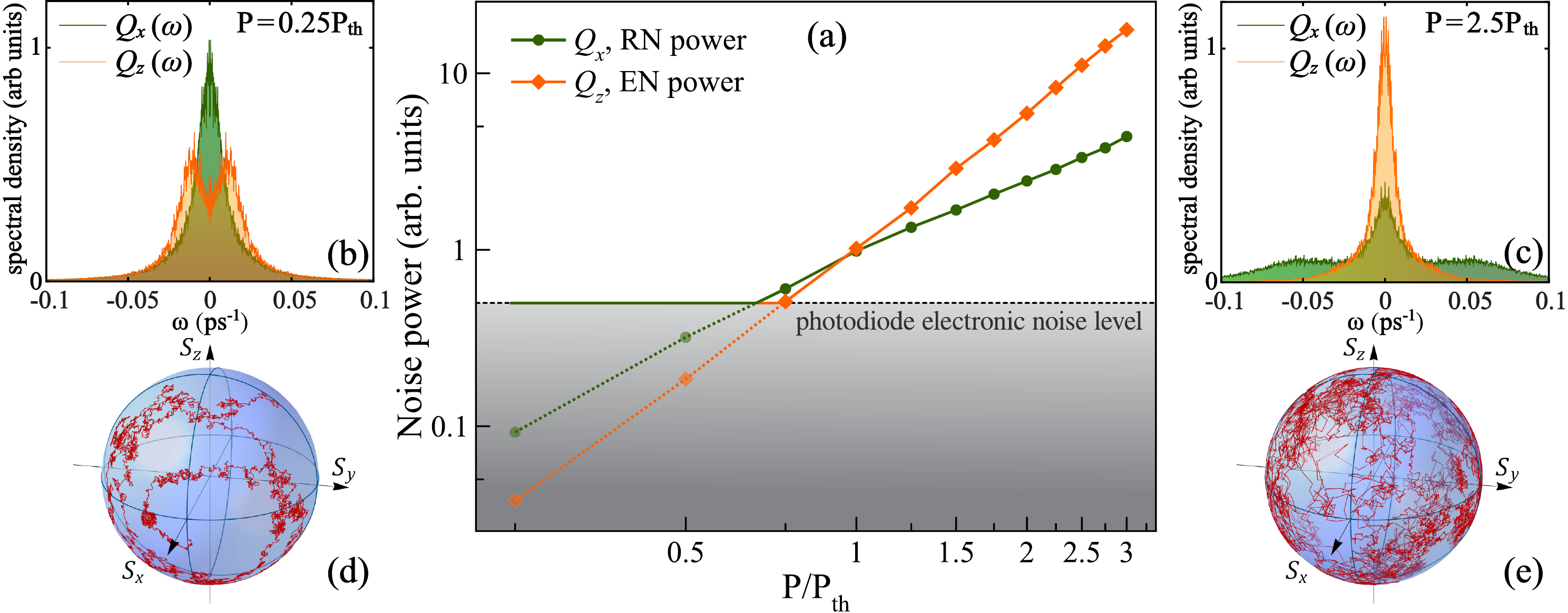}
		\caption{(Color online) Simulations of the noise dynamics of the condensate-emission polarization. The middle panel (a) shows the pump-power dependence of the zero-frequency spectral densities $Q_x(0)$ and $Q_z(0)$ correlated with the rotation and ellipticity noises, respectively. The dynamics was simulated \textcolor{black}{with noisy initial conditions} over 30 ns, and the data were averaged over 100 realizations for each point. The dashed line denotes the approximate level of the photodiode electronic noise \textcolor{black}{(shaded region)} that limits the actual range of noise power measured in the experiment, see Fig.~\ref{meas-examples}. The lower noise power values are indicated by the dotted parts of the curves. The stochastic behaviour of the polariton condensate in the subthreshold regime at $P=0.25P_{\rm th}$ is illustrated in the left panels (b) and (c) while the right panels correspond to the above threshold pump power $P=2.5P_{\rm th}$. The full-range spectral densities $Q_x(\omega)$ and $Q_z(\omega)$ are shown in (b) and (c). The corresponding dynamics of the normalized condensate pseudospin $\mathbf{S}/S$ on the Poincar\'e sphere during $1.5$~ns of evolution are shown in (d) and (e). {The parameteres used for simulations are $\alpha_1=10$~$\mu$eV, $\alpha_2=-0.1\alpha_1$, $\alpha_r=2\alpha_1$, $R=0.01$~$\rm{ps}^{-1}$, $\gamma_c=0.01$~$\rm{ps}^{-1}$, $\gamma_r=2\gamma_c$, $\delta_l=8$~$\mu$eV.}}\label{model}
\end{figure*}

Eqs.~\eqref{eq1} and~\eqref{eq2} were solved numerically on the timescales of several tens of nanoseconds, see Fig.~\ref{model}. The polarization noise properties are governed by the behaviour of the pseudospin (Stokes) vector of the condensate, whose components in the circular polarization basis read $S_x = \operatorname{\mathbb{R}e}(\Psi_-^*\Psi_+)$, $S_y = \operatorname{\mathbb{I}m}(\Psi_-^*\Psi_+)$, $S_z = \frac{1}{2} \left(|\Psi_+|^2 - |\Psi_-|^2\right)$. The stochastic dynamics predicts the width of the fluctuation spectrum to be comparable to the spectral line width, that is of tenth of THz, see Fig.~\ref{model}b,\,c. The bandwidth of experimentally used balanced photodiode was much smaller (limited by 1 MHz). Therefore the noise power measured with the balanced photoreceiver of the SNS scheme corresponds to the calculated spectral density $Q_\beta(\omega) = \int B_\beta (\tau) e^{-i \omega \tau} d\tau$ taken at zero frequency, $\omega = 0$, that is $Q_\beta(0) = \int B_\beta (\tau) d\tau.$ Here $B_\beta(\tau) = T^{-1} \int_0^T S_\beta(t) S_\beta(t+\tau)dt$\enskip is a correlation function of the quantity $S_\beta$ with $\beta$ standing for the indices $(x, y, z)$ and $T$ being the time of calculations. 

The calculated pump-intensity-dependence of the noise power, Fig.~\ref{model}c, agrees with the experimentally observed asymmetry between the subthreshold and the above-threshold regimes of the polarization noise behaviour.

The observed asymmetry can be understood examining the conservative dynamics of the Stokes vector,
$\partial_t \boldsymbol{S} \propto - \boldsymbol{\Omega} \times \boldsymbol{S}$, which corresponds to its precession about the effective magnetic field $\boldsymbol{\Omega} = (\delta_l,0,(\alpha_1-\alpha_2)S_z)$. The $x$-component of this field, $\Omega_x = \delta_l$, arises from the splitting of linear polarizations and accounts for the rotation of the condensate pseudospin $\boldsymbol{S}$ about $x$ axis. 
The field strength $\Omega_x$ does not depend on the pumping level.
In contrast, the $z$-dependent effective field $\Omega_z=(\alpha_1-\alpha_2)S_z$ does depend on the pump and appears to be suppressed at the small condensate occupancies, $S_z \rightarrow 0$, typically below the threshold.
An example of the fluctuation dynamics of the condensate pseudospin at the subthreshold pump is shown in Fig.~\ref{model}d. Due to the precession of the pseudospin $S_y$ and $S_z$ components their zero-frequency noise power turns out to be suppressed, causing the RN power $Q_x(0)$ to exceed the EN power $Q_z(0)$, see Fig.~\ref{model}b and \cite{Glazov2013}.

An increase of the pump power above the threshold results in the growth of the condensate occupancy and the increase of the effective magnetic field generated due to spin-dependent polariton-polariton interactions. By virtue of the strong spin anisotropy of the polariton-polariton interactions, which are much stronger for the same spin than for the opposite spin polaritons ($\alpha_1 \neq \alpha_2$), the fluctuations of the population imbalance between the circularly polarized condensate components $S_z$ trigger the pseudospin rotation about $z$ axis. This effect is {is a manifestation of} the pseudospin \textit{self-induced Larmor precession}~\cite{bigenwald2004}. 

Note that far above the threshold the fluctuations of the order parameter which are governed by the reservoir occupancy become so strong that the pseudospin vector spans all the Poincar\'e sphere, see. Fig.~\ref{model}e. Therefore the degree of polarization of the condensate emission approaches zero that is consistent with our experimental results.

Although the time-integrated value of $S_z$ is zero, the amplitude of its fluctuations grows with the pump power increase. Therefore, above the threshold the time-averaged modulus of $\Omega_z$ grows up and at the certain pumping intensity, which is typically close to the threshold, becomes larger than $\Omega_x$. The effective magnetic field vector tilts to $z$ axis. As a result the fast oscillations of the instant values of $S_x$ and $S_y$ components, being averaged over the time that essentially exceeds the inverse effective field strength, lead to suppression of RN power compared to the fluctuations of $S_z$, see Fig.~\ref{model}c.

This approach allows reproducing of the characteristic behaviour which is qualitatively the same for any pump beam profile. An influence of the specific shape of the pump beam can be accounted for by rescaling of the reservoir to condensate scattering rate $R \rightarrow \beta R $. The factor $\beta<1$ accounts for the spatial overlap between the condensate and the reservoir, which is dependent on the pump beam profile and is expected to decrease with the growth of the optical trap size. Our simulations show that the variation of $\beta$ alters the position of the crossing point of the EN and RN pump power dependencies.

\textit{In conclusion,}
we show that the SNS technique is capable to reveal the intrinsic { fluctuation} dynamics of { spontaneously built-up} exciton-polariton system even if the condensate emission is almost completely unpolarized. The behaviour of the polarization noise, described in terms of the ellipticity noise and the polarization plane azimuth noise, demonstrates a strong asymmetry between the subthreshold and above-threshold regimes. This asymmetry stems from the competition of the linear polarization splitting and the effect of the self-induced Larmor precession of the condensate pseudospin. In particular, we found that below the polariton lasing threshold the fluctuation dynamics is superimposed by the pseudospin precession about the effective magnetic field oriented in the equatorial plane of the Poincar\'e sphere. We ascribe the appearance of this effective magnetic field to the microcavity mirror birefringence which arises from the mechanical strain caused by the lattice mismatch of the materials. Above the condensation threshold the effect of the self-induced Larmor precession builds up as the population of circular polarization states grows up. These concurring effects lead to the intersection of the EN and RN pump-power-dependencies.

\textit{Acknowledgments.}
The work was financially supported by RFBR Grant No.~17-02-01112 which is highly acknowledged. The authors acknowledge Saint-Petersburg State University for a research Grant No.~51125686. Experiment was carried on at Saint Petersburg Resource Center ``Nanophotonics''. I.\,Y.\,C.{, P.\,G.\,S.} and A.\,V.\,K. acknowledge the support form the Westlake University (Project No. 041020100118) {and from the Program 2018R01002 funded by Leading Innovative and Entrepreneur Team Introduction Program of
Zhejiang}. A.\,T., Z.\,H. and P.\,G.\,S. are deeply thankful to Russian Science Foundation~Grants~No.~19-72-20120 for sample growth financial support. I.\,I.\,R.~appreciates the support of experimental work by RFBR grant No.~19-52-12032.

\end{document}